# Computer based Information Systems and Managers' Work.

CHRIS KIMBLE.

University of York.

and

KEVIN McLOUGHLIN.

University of Northumbria at Newcastle.

## 1. Introduction.

This paper identifies three categories of model: the Technology Impact Model; the Social Impact Model and the Integrationist Model, which imply different views of the "impact" of Information Technology on work organisation. These models are used to structure data from case studies conducted by the authors to explore the implications of the use of computer-based information systems for managers' work[*]. The paper argues that the "impact" of information systems is not a single stable and predictable outcome but a non-linear ongoing process that changes and evolves over time. It also argues that the actions of individuals and groups within an organisation are not wholly determined by outside forces: people can and do react to, and shape, systems in different ways. In this sense, the "impact" of computer-based information systems on managers' work reflects decisions made by managers themselves about how the technology is used.

---

[*] This research was funded by a grant from Northern IT Research.

## 2. Background.

The idea that the work of managers will be affected by the application of Information Technology was being discussed as early as 1958. Much of the discussion has focused on the future role of middle management and has been speculative and gloomy in its predictions[1]. Typically one of two scenarios has been advanced[2]. The first is based largely around the idea of technological determinism. Technology itself plays a key role, either leading directly to social change or acting indirectly to facilitate organisational change. The second however starts from a different viewpoint arguing that people determine the effect of a technology not the other way round.

An interesting example of these two contrasting approaches is found in the work of Leavitt and Whisler[3] and Applegate et al[4]. Leavitt and Whisler's article "Management in the 1980s" took a clearly technological determinist viewpoint arguing that the development of IT would (a) lead to top managers taking an even larger proportion of the innovating, planning and creative functions than they have now, (b) that there would be many fewer middle managers and most of those who remain would be routine technicians rather than thinkers, and (c) that IT would allow the top to control the middle just as Taylorism allowed the middle to control the bottom. In contrast the article written by Applegate et al 30 years later, argues that merely to react to new technology is a grossly inadequate response. They believe that managers should not simply respond to technological changes but should actively use them to shape the organisation. They state that the role of business leaders is to decide how to develop and use IT: they should not be driven by the technology.

The recognition of the role managers play in shaping the ways in which the technology is designed and used has prompted a more optimistic assessment of the implications of IT for managers, seeing its use as requiring new skills, freeing up more time for other valued activities such as people-management, providing better quality and more timely information to aid the decision-making process[5].

These scenarios relate to two opposing approaches to technology and social change long identifiable in the literature[6]. They are both causal or deterministic models based on the idea of one thing "impacting" on another to cause change. In the first, morally neutral technological progress impacts on the functioning of a social system. In the second, social values, expressed through the controlled and intentional application of a technology impacts on the use and design of the technology. In theory each model can predict a wide range of possible outcomes. The first however is most often associated with notions of control, predictions of job losses and de-skilling while the second is most often associated with predictions of changes that reflect the dominant social values of a group, an organisation or society[7]. A brief overview of these models, which we have labelled the Technology Impact Model and the Social Impact Model, is given below.

## 2.1 Technology Impact Model

Underlying this first model is the notion of an impartial and objective technology impacting upon its social milieu. In the Technology Impact Model information systems are seen as a substitute for labour, in much the same way as "automation" is used on the shop floor. The central argument is that technology can perform the work of managers more efficiently than a human being. Technology is usually conceived of as a machine or some technical process and is presented as the outcome of scientific progress. Technology is used to improve some mechanistic notion of "efficiency", for example, the speed or the volume of transactions processed. Typically the model is used to predict that Information Technology will lead to the deskilling of managers as their work becomes rule based and more routinised. As more integrated information systems develop the role of the human manager inevitably becomes degraded. Their role contracts progressively until eventually it disappears altogether.

## 2.2 Social Impact Model

In the second model the technology is not the cause of an impact but the agent of intentional change. It is not the technology but the way it is designed and used: subjective social values impact upon a technology and its use. The central argument is that technology does not emerge unsullied from some objective notion of scientific progress but that social values are inevitably "built into" a technology with the intention of bringing about a certain outcome. Technology is often given a broader interpretation than in the Technology Impact Model and the term may be used to include rules and procedures as well as physical entities[8]. Technology is used as a means of improving a more subjective notion of "effectiveness" such as giving people the time, the information and the organisational structures they need to take a more creative approach to their work.

Both of these models have been the subject of extensive criticism. The Technology Impact Model has been criticised for its deterministic emphasis and its view of change as involving a linear progression[9] and also its tendency to ignore the influence of human action on the development and use of a technology[10]. The Social Impact Model in turn has been criticised for pushing technology out of the picture altogether[11], for replacing one form of determinism with another, for relying too much on human agency and neglecting the social and economic forces beyond the control of the actors involved[12] and, perversely, for understating the influence of human action[13].

The incompleteness and limitations of each of these models were illustrated by an earlier study of CIM[14] carried out by one of the authors. One of the clearest themes to come from this work was the wide range of potential impacts CIM could have. Although there appeared to be some prima facie evidence to support the Technology Impact Model, on its own, it was inadequate to explain the abundance and variety of the results. The problems identified by respondents were mostly associated with being expected to manage a far reaching, but as yet undefined, change that was expected to affect every aspect of their working lives[15].

Scarbrough and Corbett[16] use the metaphor of the dance to illustrate this confused and confusing relationship between technology and the organisation. They observe that it is becoming increasingly difficult to distinguish between the dance and the dancer and that it may no longer be possible or even desirable to do so. Recently attention has begun to focus on a re-conceptualisation of technology that integrates features of both models[17]. Orlikowski, drawing on Giddens' theory of structuration[18], has developed an integrationist approach which she labels the structurational model of technology[19]. This approach is viewed as providing a means of conducting new empirical research on both the Information System development process and the implications of Information System use[20]. In assessing the value of this approach for information systems research, Walsham and Han[21] also note its value as a means of locating and re-interpreting earlier approaches such as web models and institutional analysis[22]. More recently Walsham[23] has provided a synthesis of these different approaches into a broad analytical framework designed to advance our understanding of organisational change linked to computer based systems.

## 2.3 The Integrationist Model

The Integrationist Model portrays an "impact" not as a linear outcome but as a complex, interactive and ongoing process. The principal mechanism for this is the interactions of groups and individuals free to act within the constraints of their current milieu. Note that the term "impact" is used here as a convenient shorthand for "outcome at any particular time" as, unlike the previous models, there is no real concept of an "impact" at all. The outcome at any one time both shapes future outcomes and is shaped itself by what has gone before. Technology does not "impact" on its social environment or vice versa but, over time, each shapes the other. The model can not predict an outcome in a deterministic sense, although it may be argued that a clearer understanding of what has happened in the past can help to develop a better understanding of what might happen in the future.

## 3. Evidence from Case Studies.

Having now outlined the essential features of three generic models the focus of this section of the paper will be on illustrating the empirical usefulness of these models, using data from case studies that sought to examine the implications of the use of computer-based information systems for managers' work. Throughout this set of examples the term "impact" will be used as a shorthand for "outcome of the interaction between technology and its social context at one particular time".

The primary method of data collection was in-depth, semi-structured, interviews lasting between one and a half and two hours with 65 managers from eight companies (4 in service industries and 4 in manufacturing) that had introduced integrated computer-based information systems. The precise number of interviews varied from company to company. The managers interviewed were drawn from different levels and functions within the companies. The interviews explored the views, experiences and concerns of the managers in relation to the use of information systems and their roles and responsibilities.

In addition to these interviews further interviews were held with Personnel and Information System managers to provide context and background information; short periods of observation were carried out and other documentary materials, e.g. minutes of meetings, information technology strategy plans, annual reports, organisation charts and other company publications, were also studied.

### 3.1. New Systems Shaping Changes in Existing Culture and Practice.

There was a class of "impacts" present in our work that seemed to fit most closely the Technology Impact Model: technology impacting on the organisation. The particular requirements of specific systems frequently had the effect of enforcing a more rigorous and formalised discipline upon managers. This effect was particularly obvious at middle and junior levels of management, for example, a director in Company A contrasted the old way of working (known inside Company A as "the Company One Way") with the way in which things should now be done (known as the "Company Two Way").

> "… if you going to put the lot on a machine you have to have formal procedures about who does what … you will have much more formality in 'Company Two'. The first thing you have to instil into people is the need for procedures … there are certain things that must be done and they must be done in that way and they must be done on a regular basis." (Purchasing Director, Company A.)

However it must be pointed out that, even at a superficial level, the simple model of an impartial and objective technology impacting on people soon breaks down. The introduction of these such systems was, in most cases, accompanied by an intentional shifting of accountability and responsibility down the organisational hierarchy. The same purchasing director also explained that one objective of the company's MRP II system[*] was:

> "… to force responsibility and accountability down the management line … we saw MRP II was going to enable us to do that because everyone was going to be part of this system and we could put the accountability where it really belonged." (Purchasing Director, Company A.)

The case studies provide two further examples to illustrate this class of outcome. In both of these examples the "impact" of the system was that the information it produced and distributed appeared to undermine the position of a traditionally powerful department in the company. These departments had, because of historical circumstances, enjoyed a pivotal position in the functioning of the organisation. They had built up a base of informal and unquestioned authority upon which the operation of the new system "impacted".

---

[*] MRP II (Manufacturing Resources Planning) is a suite of sophisticated programs that attempt to integrate all aspects of the planning and control of the personnel, materials and machines required to manufacture a range of products in one comprehensive computer based system.

Company A provides the first example. Company A produces chemicals for the pharmaceutical and agricultural industries. Historically their business was built up as a "Jobbing Manufacturer" of Pharmaceuticals. Pressure to increase capacity under tight budgetary constraints, among other factors, led to the decision to adopt MRP II at the site. Within Company A there had been the unquestioned belief that "production was king". People working in the production area were "the heroes" of the company who could always "deliver the goods" at the end of the month. In contrast the warehouse was seen as:

> "… something of a dumping ground. The less able, the less healthy, the older people within the company tended to migrate to the warehouse department." (Warehouse Manager, Company A.)

However with the introduction of MRP II this situation changed radically.

The company had been aware that it was carrying too much stock but the assumption had been that this was due to inefficiencies in the warehousing arrangements. In order for the new MRP system to operate effectively this situation had to change. Over time, stock location accuracy improved and a new system of requisitioning stock instituted. Once the MRP system came fully into operation the warehouse could only release material from stock if it matched the details held on the Master Production Schedule exactly. Production personnel were now required to follow a predetermined MRP Master Production Schedule and to record accurately their use of materials when and as they used them.

In the past if production requested material, the warehouse had to supply it. This practice had contributed to many of the perceived inefficiencies of the warehouse. Full drums of material would be requisitioned when half full drums were already on the site and sometimes material that had not been tested for quality was used and a whole product batch would have to be scrapped. After the system began to function production's "somewhat cavalier" attitude toward the requisitioning and recording of stock was broadcast to an audience that included senior managers and directors. Reflecting on the past behaviour of production a director described their attitude as:

> "A law unto themselves … cowboys … they were out there making the chemicals, everybody else was a hanger-on." (Director of Logistics, Company A.)

Only after the system had been introduced and begun to function did people begin to appreciate the far reaching effects it would have on the socially constructed company culture. Several managers admitted that they had unrealistic expectations of what the system could do in some areas but had overlooked the effects it could have in others. For example,

> "… with hindsight we maybe didn't … fully appreciate the ultimate ramifications of what we were going to end up with before we started … all MRP II has really done is to … highlight the shortcomings of the company which have always been there, it's just now we can actually see them." (Master Scheduler, Company A.)

The second example of this type is found in Company G. Company G is a chain of large "out of town" retail outlets. It is a family owned business built on being able to do "good deals": buying in goods cheaply and selling them on at a relatively high profit margin. A culture had developed in Company G, similar to that in Company A, where the purchasing department were the "heroes" of the company.

Company G had had POS[*] terminals for some time but previously only used them to look up the price of goods at the till. Apart from providing prices the only information they produced were sales statistics used mainly by the buyers. As part of the modernisation of the company's accounting systems the existing POS terminals were linked into a wider stock and management information system so that overheads could be more tightly controlled.

---

[*]    POS is an acronym for Point Of Sale (sometimes called EPOS for Electronic Point Of Sale). POS is a means of providing managers with the information they need to improve stock control, etc by collecting information about what has been sold at the Point Of Sale, usually by passing a product through a bar code reader.

Previously, because the system had only been used to send a price to a till, the practice had been that when there was a "special offer" on an item all that was done was to add a new item to the POS database. The result was that many product codes had been duplicated. A direct outcome of this practice was that there was no logical link between the original item and the "special offer". In order for the new system to work these duplicated codes had to be removed. The problem of removing duplicated codes was described as a "technical difficulty". However the fixing this "technical difficulty" led to far more deep seated issues surfacing. These were euphemistically described by one respondent as problems of "cultural acceptance".

Previously, like the warehouseman in Company A, the position of the store manager carried a low status. Store managers had very little direct involvement with the organisation and were described as "someone to open and shut the store". As with Company A, the introduction of the system was also accompanied by a planned shifting of responsibility and accountability down the organisational hierarchy. Part of this shift involved store managers being trained in the elements of finance so that they could handle devolved budgeting. The result was that store managers became both more accountable and more involved with the business as a whole.

When the new Information System was introduced senior managers slowly became aware that in many cases buyers were only getting a good deal by incurring a previously unseen cost. The system showed for the first time a complete picture of the stock that the company was carrying, the rate at which items were sold and, crucially, which were being put on "special offer". It became clear that the warehouses were being filled with goods that did not sell and later had to be discounted and put on "special offer".

The result was conflict between what were described as "the two big power bases within the company": the purchasing department and the retail managers.  At the root of the conflict once again was the information produced by the new system that "revealed" the practices of one group to another.

### 3.2.  Existing Culture Shaping Change Associated with New Systems.

Another category of "impacts" we observed seemed most closely to fit the Social Impact Model.  It appeared that the principal "impact" of the dominant existing culture was to undermine the stated objectives for using the technology.  As Campbell and Warner[24] observed in their study of selected British companies, if an organisation's culture demands the perpetuation of a style of working then that is what tends to happen - regardless of the technology.

The first example concerns the differential use made of the systems by senior and other levels of manager.  It was clear from the research that the way in which senior and middle managers used the systems differed considerably.  Middle managers used the systems to monitor the work of the people they supervised and to analyse and create information.  Senior managers on the other hand tended to make less use of the systems and relied upon their more junior colleagues to supply them with a paper based abstract or summary.   For example, one senior manager in Company C, a telecommunications company, commented:

> "I rarely go through the terminal because the information I get comes in on a print out … I have a filter whereby one of the managers knows the information requirements I have …" (Senior Billing Manager, Company C.)

Senior managers frequently saw the potential of information systems to remove layers of middle management from their organisations.  However their continued use of middle and junior level managers to collate information appears to work against this. The limited use of information systems by senior managers meant that they were still highly dependent on layers of middle management to supply them with paper-based information.

Senior managers put forward a variety of reasons for their low levels of use of the systems.  They admitted to a lack of skills but argued that, in any case, the systems did not contain information they wanted in a suitable form.  At first sight these appear to be essentially technical problems that could be quickly solved by better design or training.  However, a closer examination reveals that cultural rather than technical issues lie at the root of the problem.

The first issue concerns the perception of the role of a manager in relation to his or her use of Information Technology.  Several managers, both at senior and less senior levels, implied that it was somehow not part of a manager's role to use "Information Technology".  Indeed for some senior managers it appeared to be almost a measure of status that they could command human processing power, in the form of middle management or secretarial support.  Perhaps the best example to illustrate this is "The Manager as a Typist".  The phrase "The Manager as a Typist" is used to describe a belief held by many of the managers we interviewed.

This notion was held by different people with different degrees of conviction although it was more prevalent among male senior managers.  In its least extreme manifestation it simply equated the use of a keyboard with "low level clerical tasks".  In its most extreme form "The Manager as a Typist" equates the use of a keyboard directly with the use of a typewriter, and hence, secretarial work.  For example, a manager in Company E, which like Company C was also in the communications business, commented,

> "… the manager ends up as a typist. I don't see any point in paying
> managers high salaries to sit at a keyboard. If I want information I can
> simply pick up the telephone or go direct to the person concerned."
> (Senior Personnel Manager, Company E.)

Presenting such a manager with a keyboard is perceived as a devaluation of his worth to the organisation.

Clearly the belief that "computers turn managers into typists" can impact on the way the system is used by senior managers. It may also impact on the behaviour of ambitious middle managers. If the macho view that "real managers don't use keyboards" prevails, the managers who use keyboards are seen as in some way inferior. In Company D, a financial services organisation, there was at least one manager who, although he found it quicker and easier to draft his letters on a word processor, had to close his door if he did because his boss told him that "he wasn't paying him to be a glorified typist".

Another reason put forward for the low use of information systems by senior managers was the perception that "older managers cannot cope with new technologies". Older managers are widely believed to be less receptive to "new technology" and reluctant to change their working practices[25]. In most of our case studies there was a widespread assumption that older managers were unable to cope with the demands of the new technology. When restructuring was taking place it was, almost without exception, the older managers who lost their jobs.

> "… the official party line is that it's people who are not seen as having a
> role. The general characteristic is that they're over 50." (Senior Personnel
> Manager, Company C.)

Viewed superficially this might appear to be an impact of the system on the organisation however, on closer examination, this was clearly not the case.

It is all too easy for the "problem of older managers" to become a self-fulfilling prophecy. It is expected that older employees will not cope with change or new technology and so they don't. Warr[26], for example, has shown that although older employees can learn new techniques effectively they feel less confident in their abilities and receive less formal training than their younger colleges. Similarly, our case studies provided a number of examples to suggest that older managers do not necessarily find learning new systems any more difficult than some younger managers.

> "… a lot of reservations from older people, but, in the end, we never had one failure of (the 250 people in the division) … to come to terms with the system." (District Billing Manager, Company C.)

The belief that "older managers cannot cope with new technology" appeared to provide what Zuboff[27] called a "ritual justification" for making older managers redundant and provides another example of how a socially constructed doctrine can shape the impact of a new system in a way that appears to undermine stated objectives.

Among the reasons for the introduction of the new systems in our case studies was the desire to make the company "fit to survive into the twenty first century". However the gain of a lower wage bill and less complex industrial relations needs to be weighed against the cost of losing valuable management knowledge and expertise. In the long term using IT as a justification for losing older managers creates a potentially more serious problem. Between now and the year 2010 there will be a decline of 20% in the number of those aged between 20 and 30 in the UK and a corresponding increase in the number of 40 to 60 year olds. A planning manager in Company A commented:

> "We've had a policy of non-recruitments and early retirements … I have this recurring dream that … come the year 2015 on one Friday afternoon everybody in the company's going to retire and on the Monday morning they'll … open the gates but nobody will come through because nobody's employed any more." (Planning Manager, Company A)

Company A provides one final example to illustrate that cultural influences can shape the impact of a system. This example illustrates the "impact" of external, rather than internal, culture. Company A is a division of an American owned multinational company. Many of the senior managers in Company A were constantly making the point that the new MRP system required people to abandon the old way of doing things and to stick to the schedule produced by the system. This transition was obviously a difficult one to make for a number of reasons, some of which have been outlined above. The problem was exacerbated however by the culture of the parent company who had dictated the form the information system should take in the first place. One director confided:

> "I don't think our lords and masters in America have yet made the mental jump because they still are using the old criteria for measuring performance, they still think the old way, and it's taking time to educate them." (Purchasing Director, Company A.)

Another manager explained:

> "… we've got a very demanding lady president, lady might be the wrong terminology, but we've got an American lady who basically runs the chemical group, and she has a habit of ringing up the MD to find out how we're doing against a particular order …" (Master Scheduler, Company A.)

The result of the phone call is that the schedule is changed. Again the effect of this external cultural influence is to shape the "impact" of the system and to undermine the way in which the system is supposed to function.

### 3.3. The Role of Interaction and Choice in Shaping Outcomes.

In the first set of examples it was largely the features of the technology itself that shaped its "impact". In the above examples existing culture and relationships appear to play a dominant role in shaping the "impact" on the organisation. The final set of examples illustrates how the interaction of people and technology shape an outcome over time. These examples fit our third model most closely.

It is important to view manager's use of these systems within their own conception of their wider role. Throughout the research, managers stressed that the use of these systems was only one part of their job. The information the systems provide may be useful, even essential for the manager's work but, in the end, managers see their use of the system as being of only marginal importance to their job as a whole. A manager in Company B, a manufacturing division of a British owned multinational, commented

> "… it's still only a tool to enable management to make a better decision. It gives you more information … it saves a lot of legwork but at the end of the day all it says to you [is] that's the information … then it's down to me personally …" (Planning Manager, Company B.)

The essence a manager's use of the system is in how the information it produces is used - rather than how it is obtained.

The clearest illustration of this is found in the different approaches taken by two managers in the same organisation, Company C. The first manager provided this example of how he used the system to monitor the work of his subordinates. In answer to the question "Does the system enable you to exercise more control over your subordinates?" He answered:

> "Yes definitely … what that means that is a very tight control on a level 1 manager. He's having to report to me weekly on the number of jobs he's done in that category and where they went wrong … I'm in here at 7.05 and they're still in bed. This morning at 7.10 I knew how we'd done yesterday, and they know that I know, because, as soon as I see them, I say, what happened here?" (Repair Manager, Company C.)

The second manager, in the same department of the same organisation, uses the same system but answered the question in a totally different way.

> "I can actually see the work they're doing electronically … the people that work under you aren't under pressure (because) the boss is sitting beside them … having said that I do sit with them … you could possibly do the whole job electronically but at the end of the day … they're your people and you've got to be seen to be out there with them." (Repair Manager, Company C.)

Taken in isolation the first might have been used as an example of the system impacting on the working practices of managers. The system provides information on people's work loads and working patterns automatically and so managers become divorced from day to day contact with the people they manage. Similarly the second example, taken in isolation, might have been used as an example of how the culture of an organisation shapes the way that the system is used. The culture is one of "the personal touch" and so that is what happens even although a more efficient technology exists. In this case however both approaches exist in the same organisation at the same time. Two individual managers use the same system but in entirely different ways. It was not the system, or the dominant culture of the organisation, that determined the "impact" but the choice made by the individual manager about how to use the information provided by the system.

The ability of groups and individual managers to shape the "impact" of Information Technology plays a central role in the way in which the "impact" of the technology changes over time. The longest case study ran over a two year period. It was clear that, even over this relatively short period, the "impact" a system has may change considerably as groups or individuals react to the changes in their circumstances. Company A provides a clear example to illustrate this point.

When MRP was introduced, it quickly became apparent that up to 70% of the processes were failing to meet their target dates. The production schedule was being constantly altered and potential problems deferred until the following month. This was termed "The Bow Wave Effect". A plan was produced in the first week of a month. By the end of the second week production had looked at the plan and, with the consent of the planning department, altered the parts of it they claimed were not feasible. This happened every month. Part of one month's plan was always being carried forward into the next month. The manager in overall charge of implementing the system commented:

> "We understood the mechanics of this bow wave effect and we thought we'd cracked it … So what we decided to do, wrongly as it turned out, was to say … the capacity in the plant is not as big as we think it is. So in any 30 day period we've got to expect to lose say 3 days through untoward events that are not modelled." (Head of Project, Company A.)

Production managers accepted the revised schedule but still ended up reducing it slightly. The conclusion that the senior managers reached was that:

> "… plant managers are simply using this … figure as the target … all we've done by taking days out is we've given them a lower target." (Head of Project, Company A.)

This view was confirmed by the observations of the warehouse manager.

> "It's an awful thing to say but I've just started to realise (that) production actually delay schedule one day but then during the night they'll (say) I'm ahead of myself … They've delayed it the first day but they're bringing it back on-line the second day … They're gradually getting to know … how they can help themselves." (Warehouse Manager, Company A.)

Clearly the initial effect of the MRP system was to change the role of production dramatically. The old culture where "production was king" and they were "a law unto themselves" had, it seemed, been fundamentally changed. In the company's own terminology 'Company Two' was in the ascendancy. It would however be naive to think that such a cultural shift could be made overnight. Over a period of time the features of the new system and the existing company culture interacted to shape a new outcome where certain features of 'Company One' began to reassert themselves. This new outcome was shaped both by features of the technology and by features of the corporate culture but was mediated through the actions of individuals and groups with a common interest.

The "impact" of these systems is not a one off effect but something that evolves over time. The time it took for systems to "bed in" often appeared to have been completely unexpected. For example, some fourteen months after the change over to MRP II, the manager responsible for its implementation in Company A commented:

> "… I guess I naively thought that maybe within a year or so after putting it in we would have everything sorted. I think now we are probably talking about 2 to 3 years." (Head of Project, Company A.)

A senior manager in Company C made a similar observation. He commented that, once the new system had been introduced, many members of his staff felt "that was that" and their work could "go back to normal". He observed that he did not have any problems with the initial introduction of the systems because:

> "… staff are actually keen when they have got something new turning up. There is the worry of the new technology to get over, which is very hard, but at the same time they are receptive to change". (District Billing Manager, Company C.)

However once the new system was in place he began to find resentment whenever changes were needed to "fine tune" the system.

> "… fine tuning in many ways becomes the most difficult because folk say 'hey, I'm happy with this, why do you have to change that' and 'huh more change!' … it's when you think you've succeeded that your real problems come in because the world doesn't stop." (District Billing Manager, Company C.)

It was clear from the case studies that individuals and groups could, and did, shape the "impact" of systems in ways that made the idea of some form of generic impact model a nonsense. It was clear that what ever form these "impacts" might take they did not in any sense represent a final outcome: change was a continual and ongoing process.

## 4. Discussion

This paper has introduced several examples to show how social processes can shape and be shaped by information systems. It clearly demonstrates that the richness and complexity of outcomes make the Technology Impact Model and The Social Impact Model, which both predict a non existent uniformity of outcome, inadequate as tools for analysis. The material from the case studies however does show that "impacts" ostensibly similar to models one and two do seem to occur.

Our first set of examples illustrate how an intrinsic feature of an information system, its ability to collate and distribute data, can disturb the status quo in an organisation. In character this class of observations is superficially similar to our Technology Impact Model. The introduction of these systems required managers to follow a set of more structured procedures and, in one sense, a dimension of managerial freedom of action was constrained. However, as we saw later, managers see their use of the system as being of only marginal importance to their job as a whole: the essence of manager's work is in how the information is used rather than how it is obtained.

The most immediate "impact" for most managers in this first set of examples was a change in the perceived status of groups within the company that was related to the way the system portrayed their performance. The information produced and distributed by the new systems disrupted unquestioned, and possibly unrecognised, informal relationships between departments. What Keen[28] calls "The Politics of Data" clearly had a role to play. A "new reality" for the organisation was created based largely on the information created, stored and distributed by the system.

This however was not an illustration of a simple impact of an objective technology on an unsuspecting organisation. The apparent "impact" of the systems on the relative status of different groups in the company was not solely an "impact" of the system. The decision to train store managers, or the decision not to allow production personnel to control the withdrawal of materials from the warehouse, was central in shaping what might otherwise appear to be have been an impact of a technology on an organisation.

Should we then interpret this observation as a variant of our second model where the hidden social goal is to increase managerial control? Were the managers in our case studies Machiavellian Social Engineers subtly altering their employees sense of reality by the choice of what data to record? As with the earlier work on CIM[29] there appeared to be little evidence to support this view in the organisations we studied. Most managers did not appear to appreciate the political dimension to data fully, if at all, until some time after the system was in place. Senior managers in particular did not appear to have a particularly sophisticated view of what the systems could do at this level or how they might be used. It is the author's contention that these observations can be most convincingly interpreted by an application of the third class of models: The Integrationist Model.

A similar approach can be taken to the second set of examples. Again, at first sight, this class of impacts appears to fit most closely to the second model. Preconceived notions of the abilities and motivations of older managers, or views about the appropriateness of managers using keyboards, provide examples of how socially constructed doctrines can shape the "impact" of systems. These "impacts" are not due to features of the system but result from the value systems of the managers themselves.

The actions taken by the managers who hold these beliefs however appear to undermine some of the reasons for introducing the technology in the first place. Several senior managers, for example, expressed the belief that these new systems would facilitate de-layering and enable a redistribution of responsibility and accountability. However, by shunning the use of the technology themselves, and by continuing to rely upon juniors to collect and collate information for them, their actions mitigated any of the effects the systems might have had in this area. Should we therefore depict cultural values and social action not as the driver of technology but as a "barrier to progress"?

Once again, we would argue that viewing this set of examples as a corrupted form of the Social Impact Model is to oversimplify the situation. This apparent "impact" of social values is not an end in itself but part of an on-going process: an outcome that evolves over time as a result of the continual shaping and re-shaping of both the technology and its social context.

The "impact" of the systems changed subtly over the period of the case studies. This effect was particularly noticeable in the larger case studies that ran over a longer period. In some cases managers who had initially felt that using a keyboard was somehow inappropriate changed their minds. In other cases managers who initially felt that the systems had led to a loss of face to face contact found that they had opened new channels of communication. Although organisational culture clearly played an important role in the way the system was used it was also clear that, over time, the existence of the new technology also played a part in shaping culture. Again there can be little doubt that simple deterministic arguments that "technology shapes organisations" or "organisations shape technology" are incomplete.

If the key to interpreting these seemingly contradictory findings lies in the application of the third model what features of this model has this work highlighted? Two features of our findings seem to be of particular importance. The first is the recognition that the "impact" of these systems is not a single static outcome but an ongoing process that changes and evolves over time. The second is that individuals and groups within the same organisation can react to, and shape, the systems in different ways.

The statement that the nature and direction of the "impacts" changes over time is, at one level, stating the obvious. Over a period of years it is only to be expected that both the technology and the environment in which the company operates will change. Such changes must inevitably affect both the system and the way it is used. In Company A for example, the system was three years in the discussion stage and it was expected to take at least three years to bed in. In this time Company A short listed three different MRP systems and changed ownership once. Similarly at Company G there were dramatic changes at board level during the two years that their system was in development.

However, even if a technology, an organisation's environment and the company culture do not change significantly, it is still to be expected that, the impact of these systems will change over time as people learn to accommodate to their changed circumstances. In Company A for example, production managers began to find ways of "helping themselves". In turn senior management reacted by changing the way the system presented production with a schedule of work. In Company C there were similar examples of how, when lower level managers found ways of disguising what was happening in their area from the superiors it was decided to change the method of supervision.

## 5. Conclusion.

The underlying objective of this work was to gain empirical insights into the "impact" of information systems. Three interpretative models have been described, data collected and analysed, and insights gained. What conclusions can be drawn and what recommendations can be made?

The two most significant conclusions to be drawn from the data concern the recursive nature of the systems development process and the role of human actions in shaping the outcome in any particular instance. Although the data from the case studies can be interpreted using any of the three models so called "impact" models are inadequate to explain the richness and diversity of the data. A central conclusion of this work must be that the term "impact" can only be sensibly interpreted within the framework of a model that recognises both the indeterminacy of any particular outcome and the ability of groups or individuals to shape their own "impacts" over time.

The paper argues that the third integrationist model emerges as, theoretically and practically, the most useful for advancing study of the implications of computer-based information systems on managers' work. The key conceptual focus of this model is the way an "impact" develops over time. This feature is emphasised in our research findings and demonstrates the importance of a processual approach to the study of the "impact" of information systems.

The authors believe that theories and methodologies that attempt to locate change in its social and historical context should be developed further. However we argue against adopting perspectives that focus only on temporally specific instances of technological or organisational change. Instead our research evidence provides further support for the need to develop an integrationist approach that seeks to conceptualise the links between context, process and human action and which highlights the mechanisms through which such "impacts" evolve.

The practical implications for those involved in managing the change process, and for managers in general, is that technological and organisational development needs to be viewed as a continual and ongoing process. We found in our research that managers were not always prepared for the long time scale over which change takes place, nor for the many twists and turns of the change process. Adopting this view will enable the identification a wider set of factors which are relevent to the effective operation of such information systems. These factors will not lie in the technology alone, nor solely in the nature of the social values that shape the context in which it is developed, but will also rest upon how the managers themselves choose to use the technology.